# On Formal Reasoning on the Semantics of PLC using Coq

Jan Olaf BLECH and Sidi OULD BIHA

fortiss GmbH, Munich         INRIA, Tsinghua University

Programmable Logic Controllers (PLC) and its programming standard IEC 61131-3 are widely used in embedded systems for the industrial automation domain. We propose a framework for the formal treatment of PLC based on the IEC 61131-3 standard. A PLC system description typically combines code written in different languages that are defined in IEC 61131-3. For the top-level specification we regard the Sequential Function Charts (SFC) language, a graphical high-level language that allows to describe the main control-flow of the system. In addition to this, we describe the Instruction List (IL) language – an assembly like language – and two other graphical languages: Ladder Diagrams (LD) and Function Block Diagrams (FBD). IL, LD, and FBD are used to describe more low level structures of a PLC. We formalize the semantics of these languages and describe and prove relations between them. Formalization and associated proofs are carried out using the proof assistant Coq. In addition to this, we present work on a tool for automatically generating SFC representations from a graphical description – the IL and LD languages can be handled in Coq directly – and its usage for verification purposes. We sketch possible usages of our formal framework, and present an example application for a PLC in a project demonstrator and prove safety properties.

## 1 Introduction

Discovering and validating properties of Programmable Logic Controllers (PLC), is a prerequisite for the development of safety critical embedded systems in industrial automation. Tools and techniques for different kinds of systems and analysis scenarios have been developed. These comprise techniques aimed for distinct usage scenarios based on model checking and abstract interpretation.

In this work, we describe a general purpose way for the verification of PLC that are modeled using the Instruction List (IL), Sequential Function Chart (SFC), Ladder Diagram (LD), and to some extend Function Block Diagram (FBD) languages of the IEC 61131–3 [23] standard. The standard is mainly used for modeling PLC functionality in the development of embedded systems for the industrial automation domain. We describe the semantics, proving principles and an associated tool. The tool generates for a given PLC description provided in the graphical SFC language automatically a Coq [14] description and some basic theorems and their proofs. In addition to the SFC language, text based IL programs and textual representations of LD programs are used in our PLC descriptions. We have formalized syntactic representations of IL and LD. Thus, programs given in these languages can be imported directly into our Coq environment. Furthermore, we describe a high-level formalization of FBD. We present some standard techniques to reason about our PLC descriptions and verify properties. Furthermore, we sketch a general usage scenario and apply our techniques to a case study on a PLC used inside a sorting machine.

The formalization of the IL, SFC, LD, and FBD semantics is done in the formal proof system Coq and its extension SSReflect [16]. Choosing Coq enables us to use its extraction mechanisms later and produce a certified compiler or interpreter for PLC based on our semantics. In this development, we also use some SSReflect libraries. In particular we use the libraries on booleans, natural numbers, lists and generic interface for types with decidable equality.

Thus, the most important contributions of this paper comprise:

- Formal Coq semantics and connected proofs of the IL, LD, SFC languages and a high-level formalization of an FBD semantics which are reusable for other projects.

- A tool to automatically generate SFC representations and some properties and proofs.

- A method for the verification of PLC properties of our Coq descriptions of a PLC that has been established based on our verification experiences.

The work presented in this paper summarizes, continues and builds upon previously published work [9, 22, 6, 5].

**Overview**

This report is organized as follows. We give an overview of PLC systems in Section 2. The main characteristics of our formalization and possible usages are described in Section 3. The IL and SFC language are presented respectively in Section 4 and Section 5. Additional languages and a transformation to IL is described in Section 7. A tool for generating Coq readable SFC representations and related proofs is described in Section 6. Example usages and an evaluation is presented in Section 8. Section 9 discusses related work and a conclusion is featured in Section 10.

## 2 Programmable Logic Controller

A PLC is composed of a microprocessor, a memory, input and output devices where signals can be received from sensors or switches and sent to actuators. Figure 1 shows the architecture of a PLC system. A main characteristic of PLC is their execution mode. A PLC program is typically executed in a permanent loop. PLC program execution can be structured into *scan cycles* which are associated with a cycle time, the inputs are read, the program instructions are executed and the outputs are updated. The cycle time is often fixed or has an upper bound limit. Therefore the instructions which are scheduled to be executed in the cycle should terminate during the cycle time interval. PLC like systems, e.g., realized using PC components follow a similar scheme and are often also programmed using the IEC 61131-3standard.

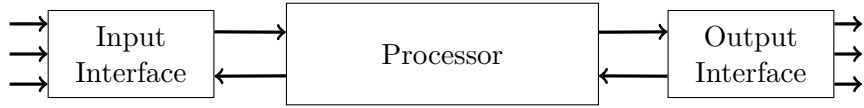

Figure 1: PLC system

**Programming languages**  Since the introduction of PLC in the industry, each manufacturer has developed its own PLC programming languages. In 1993, the *International Electrotechnical Committee* (IEC) published the *IEC 61131 International Standard* for PLC [23]. The third volume of this standard defines the programming languages for PLC. It defines 5 languages:

- *Ladder Diagrams* (LD) : a graphical language that represent PLC programs as relay logic diagrams.

- *Functional Block Diagrams* (FBD) : a graphical language that represent PLC programs as connection of different function blocks.

- *Instruction List* (IL) : an assembly like language.

- *Structured Text* (ST) : a textual (PASCAL like) programming language.

- *Sequential Function Charts* (SFC): a graphical language for describing top-level control-flow and associated data-flow in the PLC.

The last language differs from the other. It corresponds to a graphical method for structuring programs and allows to describe the system as a parallel state transition diagram thereby specifying the overall control flow of the PLC. Each state is associated to some actions. An action is described using one of the other 4 PLC programming languages like. SFC are well suited for writing concurrent control programs. In this paper we concentrate on the LD , IL and SFC languages, and to a minor extend on the FBD language.

## 3 Formalization of PLC

In this section we present a general overview on our framework and possible usages. We explain our framework in its current status and show possible benefits from future

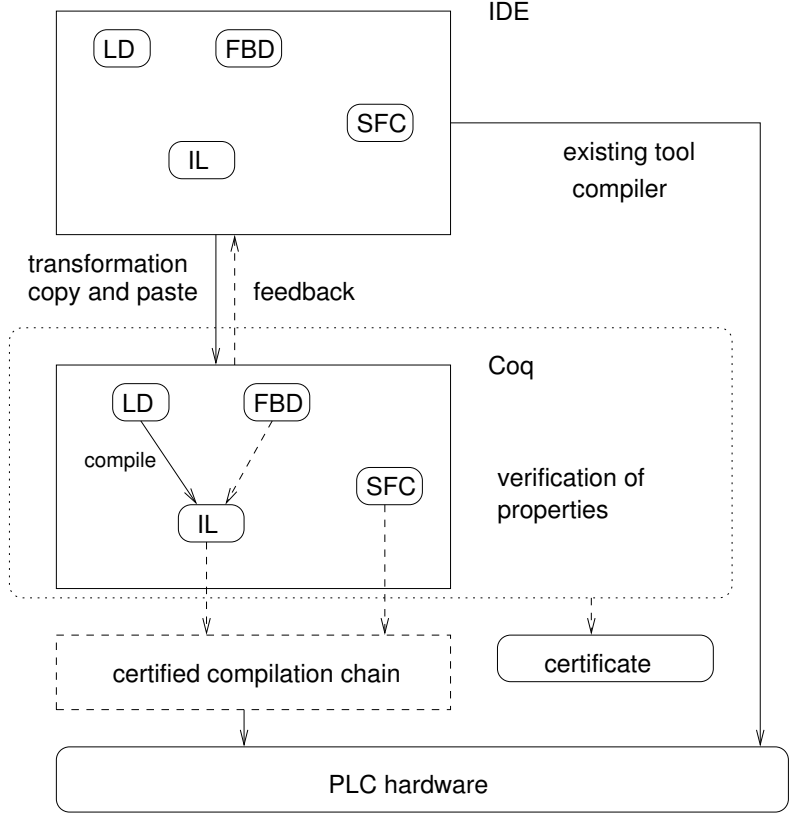

Figure 2: Our Tool Chain

extensions and connections to related projects.

### 3.1 Overview on the Framework

We present here the general architecture of our formal development for the verification of PLC programs. Figure 2 shows the tool chain we are aiming at. The IL, LD, SFC and to some extend FBD languages are used to model a system. An existing tool (for SFC we use EasyLab [3]) and compiler is there to compile these languages first to C code, then to machine-code of the PLC hardware.

In this report, we concentrate on IL, LD and SFC code and only to a minor extend on the more ambiguous FBD code. In order to carry out our verification work, we need to transform this code into a Coq representation first. This transformation into a Coq representation comprises the following steps:

- IL code can be imported directly as Coq object by using inductive types.

- For the graphical SFC language, we present a tool that transforms SFC files into Coq representations automatically: CertPLC. CertPLC also generates auxillary definitions and proofs.

- The graphical LD code can be transformed relatively easily into a syntactical representation manually. Here, we present a verified translation from LD into IL code for further reasoning and processing.

Once we have Coq representations of systems specified in these languages, we can already achieve verification results using these representations and their semantics. These results comprise the verification of various properties. Proof attempts can be used as feedback in the development and completed verification results can be encapsulated with the Coq representation to form an independantly distributable and checkable certificate. Another aspect that is shown in the figure is a long term goal which has not been achieved yet: We are aiming to connect the tool chain to a certified compilation chain (in a style as described, e.g., in [18]) and compile IL and SFC code as represented in Coq into machine code of the PLC hardware. For the compilation of LD we present a transformation from LD to IL in this paper.

### 3.2 Using our Framework to Create a Coq Certificate

The framework presented in this paper is developed to support the following standard usage scenario which aims at creating a PLC specification, its Coq representation together with proofs for properties. We call all these Coq files together a certificate. It encapsulates all necessary information for guaranteeing that the stated properties do indeed hold.

- A PLC is developed using the following work-flow:
    1. Establishing requirements,
    2. and derive some early formal specification of these requirements.
    3. Based on this specification the overall structure – e.g., the control flow – is specified using the SFC language. More fine-grained behavioral aspects are textually specified, e.g., by annotating the SFC structure.
    4. Taking the requirements and this specification, developers potentially using the help of automatic verification tools derive and specify consistency conditions and properties that must hold on the PLC. Some consistency conditions may directly correspond to a subset of the requirements.

- Regarding 3) the SFC structure may be modeled a graphical tool: EasyLab [3] or imported into this tool. The use of EasyLab allows the invocation of CertPLC to generate a Coq representation of the SFC structure.

- Regarding 4) properties are transformed to the Coq syntax by trained developers. It is not required to do any proofs in Coq at this stage.

- CertPLC may be used to generate representations for the SFC part of the PLC specification and basic proofs.

- The PLC development is further refined and fine grained parts are implemented using IL, LD, FBD languages from the IEC 61131–3 standard. Regarding Coq: The textual IL code may be imported into the CertPLC generated SFC code directly, LD code has to be written in the Coq syntax while potential FBD parts have to be formalized in our semantics framework individually.

  At this stage, we have a complete IEC 61131–3 specification of our PLC and a complete Coq representation of it together with some basic proofs generated by CertPLC.

- Properties of the PLC may now be proven interactively, e.g., by using special tactics for SFC. These hand-written parts may use the automatically generated basic proofs.

Feedback, e.g., the fact that one is not able to prove a desired property, including some counterexamples may now be used to refine the specification and go through the steps again. Once one is content, the proofs can be used to create a certificate. The certificate can be distributed and analyzed independently by third parties. One overall goal is to convince certification authorities and potential customers of the correctness of PLC with the help of certificates. Since the certificates are independent of the original development and its tools some confidential data (e.g., the certificate generation mechanism and the analysis algorithms used to discover properties of the system) does not have to be revealed during the process of convincing customers or certification authorities.

The described usage scenario can be adapted. It is, e.g., possible to integrate further hand written specifications and proofs.

## 4 Instruction Lists

IL is an assembly like language used for programming PLC systems. An IL program example is the following:

| LABEL | OPERATOR | OPERAND |
|-------|----------|---------|
| l1:   | LD       | x       |
|       | AND      | y       |
|       | LD       | z       |
|       | ORs      |         |
|       | JMPC     | l1      |

In the first line of the example above, the value of the variable x is loaded on a stack. After the execution of the second line, the stack contains the conjunction of x and y. In the third line, the value of z is put on the top of the stack. The instruction ORs of the example above removes the two previous values loaded on the stack and replace them with $(x \wedge y) \vee z$. The branching instruction JMPC is executed if the value at the top of the stack is equal to *true*.

We consider here a significant subset of the IL language defined by the IEC 61131-3. This subset covers assignments instructions and boolean and integer operations. It covers also comparison and branching instructions and *on-delay timers*. We choose to consider only booleans and integers as basic data types. In most of PLC systems, floating-point numbers are available as basic data types, but rarely used. In practice, floating-point numbers computation costs much time and may be delegated to an external device that can communicate with the PLC. This is motivated by the need to keep the program scan cycle within a relatively small time upper bound. The IL model we present in the following is an extension of the model defined in previous works [22, 9].

## 4.1 Syntax

An IL program is a variable declarations followed by a list of instructions. An IL instruction starts with an operator that can be followed by one or more operands: variables or constants. In an instruction, the operator can be preceded by a label.

Constants:
$$cst \quad ::= \quad n \in \mathbb{Z} \mid b \in \mathbb{B} \quad \text{integer or boolean literal}$$

Operands:
$$op \quad ::= \quad id \mid cst \quad \text{variable identifier or constant}$$

Instructions:

$$
\begin{array}{rll}
i \quad ::= & \texttt{ST } id \mid \texttt{STN } id \mid \texttt{SR } id \mid \texttt{RS } id & \text{store, set and reset} \\
\mid & \texttt{LD } op \mid \texttt{LDN } op \mid \texttt{LPS} & \text{load operations} \\
\mid & lb \; : \; \mid \texttt{JMP } lb \mid \texttt{JMPC } lb \mid \texttt{JMPC } lb & \text{label and jump} \\
\mid & \texttt{AND } op \mid \texttt{OR } op \mid \texttt{XOR } op & \text{booleans operations} \\
\mid & \texttt{ANDN } op \mid \texttt{ORN } op \mid \texttt{XORN } op & \\
\mid & \texttt{ANDs} \mid \texttt{ORs} \mid \texttt{XORs} & \\
\mid & \texttt{ADD } op \mid \texttt{MUL } op \mid \texttt{SUB } op & \text{integer operations} \\
\mid & \texttt{ADDs} \mid \texttt{MULs} \mid \texttt{SUBs} & \\
\mid & \texttt{GT } op \mid \texttt{GE } op \mid \texttt{EQ } op & \text{comparison} \\
\mid & \texttt{GTs} \mid \texttt{GEs} \mid \texttt{EQs} & \\
\mid & \texttt{TON } id \; , \; n & \text{on-delay timer} \\
\mid & \texttt{RET} & \text{end of program}
\end{array}
$$

The data domains of IL constants is the union of integers $\mathbb{Z}$ and booleans $\mathbb{B}$. In practice integers used in PLC are bounded. For simplicity, we restrict ourselves to unbounded integers in the presentation of this work. The integration of the support for bounded integer can be done in a modular way.

In the following, we denote the type of IL instructions by *Instr* and the type of IL *code* by *ILCode*. An IL *code* is a list of IL instructions or objects of the type *Instr*.

**Difference with the standard**  The IL syntax we defined above is slightly different from the one defined by IEC 61131-3. The main difference is the use of a stack for the evaluation of temporary results. In the standard, a special register is used for this purpose. For operators with two arguments, like the disjunction OR, parenthesis are used to indicate that evaluation of the operator shall be deferred. Figure 3 shows an example of IL program written in the IEC 61131-3 syntax (Figure 3(a)) and in the syntax we defined above (Figure 3(b)). This choice is motivated by the fact that most of the PLC manufacturers do not use the parenthesis convention of the standard. They instead use a special operator without operands that will use the stack to read his two arguments (like the `ORs` operator).

```
LD    X        LD    X
OR(   Y        LD    Y
AND   Z        AND   Z
)              ORs
ST    R        ST    R
   (a)            (b)
```

Figure 3: Differences with the standard

Another difference with the standard is the use of the operator `LPS`. This operator duplicate the value on top of the stack. It is used to keep in the stack a result that will be consumed by an operation like the one that store a result in a variable. In practice it is also used in the translation of an LD rung with parallel rears into an IL program. We will come to this later in the section on the translation from LD to IL.

**Timers** In the context of PLC applications, there is often the need to control time. For example, a motor might need to be activated or switched off for a particular time interval. Another example, in a chemical plant a valve is open and a tank will be full after a period of time. PLC timers are components that set on a boolean output after or for a period of time following the activation of a boolean input. They are used to control output signal duration or as input signal for time dependents PLC programs. In general, they have two inputs and two outputs. Figure 4 shows the IEC 61131-3 graphical representation of timers. In this representation, `IN` and `Q` are respectively the boolean input and output of the timer. `PT` is the constant input used to specify the time delay of the timer. `ET` is the output indicating the elapsed time since the activation of the timer. The delay `PT` and elapsed time `ET` are multiples of a system predefined time base.

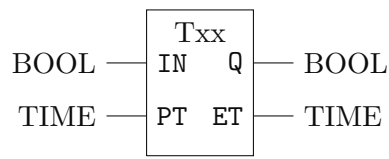

Figure 4: Standard timer representation

In our IL model, we make an abstraction of how timers are defined and see them as a *black box*. This is done by using the operator `TON` that have three arguments and uses also the value on top of the stack. In real PLC, timers are predefined function or macro that uses a global time clock. Since in our model we do not support IL function call, we choose to make abstraction of how timers are defined and use a special command to simulate the call for timers.

## 4.2 Semantics

We defined a small step operational semantics of IL programs. For the purpose of modeling PLC timers, we suppose having a global discrete time clock and that each program execution cycle has a fixed time duration denoted $\delta$. In real PLC system, $\delta$ is a global constant.

**Stack** We use an evaluation *stack* for the current result computation and also to store intermediate results that will be pulled back when an instruction like `ADDs` or `ANDs` are executed. A *stack* $V := v_1, \ldots, v_m$ is a finite sequence of data values. In the following we use the standard stack operations *push* (add an element to the stack), *pop* and *pop2* (remove respectively the top and the two top elements of the stack), *top* and *top2* (return respectively the top and the two top elements of the stack).

**States** Functions from variable identifiers to data values. They represent the program variable states and are denoted $\sigma$ of the type $\mathcal{S} = Var \to D$, where $D$ is the union of the IL variables data domains: $\mathbb{Z} \cup \mathbb{B}$.

**Configurations** Elements of the set $\mathcal{E} = Stack \times \mathcal{S} \times ILCode$. A configuration $(s, \sigma, c)$ corresponds to a stack $s$, a state $\sigma$ and a code sequence $c$ representing the current position in the program as in the LD semantics defined in Section 7.2.2. In a previous definition of IL configuration [22], we add two components that will identify the current cycle and the mode of the execution. That components are useful when we want to prove some properties over the execution of an IL program. But for the translation from LD programs, they are not needed. So here we define a configuration using the state of the stack and variables and the *continuation* to encode the program counter or position.

**Transitions** Relation on configurations $\subseteq \mathcal{E} \times \mathcal{E}$. Figure 5 gives some relevant inference rules of the IL configurations transition relation. The transition relation is also defined by an initial configuration $(s_0, \sigma_0, 0)$, where $s_0$ is the empty stack and $\sigma_0$ is the initial state that maps all the integer variables to 0 and boolean variables to $false$.

The first four transition rules of Figure 5 correspond to the *load* and *store* instructions. In the first case the stack is updated while in the second the variable state is updated. The transitions corresponding to the *set/reset* instructions (rules SR and RS) update the variable state function with the corresponding values for the given operands and the top of the stack.

The rule for the `LPS` operator modifies the stack of the corresponding configuration by *pushing* to its top a copy of its top value. The two top objects of the new stack will have the same value. In the inference rule JMP, transition for the unconditional branching

$$\text{LD} \; \frac{s' = \text{push } op \; s}{p \vdash (s, \sigma, [\mathbf{LD}\; op].c) \to (s', \sigma, c)} \qquad \frac{s' = \text{push } \neg op \; s}{p \vdash (s, \sigma, [\mathbf{LDN}\; op].c) \to (s', \sigma, c)} \; \text{LDN}$$

$$\text{ST} \; \frac{s' = \text{pop } s \quad \sigma' = \sigma[x \mapsto \text{top } s]}{p \vdash (s, \sigma, [\mathbf{ST}\; x].c) \to (s', \sigma', c)} \qquad \frac{s' = \text{pop } s \quad \sigma' = \sigma[x \mapsto \neg \text{top } s]}{p \vdash (s, \sigma, [\mathbf{STN}\; x].c) \to (s', \sigma', c)} \; \text{STN}$$

$$\text{SR} \; \frac{\begin{array}{c} s' = \text{pop } s \\ \sigma' = \sigma[x \mapsto \sigma(x) \vee \text{top } s] \end{array}}{p \vdash (s, \sigma, [\mathbf{SR}\; x].c) \to (s', \sigma', c)} \qquad \frac{\begin{array}{c} s' = \text{pop } s \\ \sigma' = \sigma[x \mapsto \sigma(x) \wedge \neg \text{top } s] \end{array}}{p \vdash (s, \sigma, [\mathbf{RS}\; x].c) \to (s', \sigma', c)} \; \text{RS}$$

$$\text{LPS} \; \frac{s' = \text{push } (\text{top } s) \; s}{p \vdash (s, \sigma, [\mathbf{LPS}].c) \to (s', \sigma, c)} \qquad \frac{c' = \texttt{findlabel}(p, l)}{p \vdash (s, \sigma, [\mathbf{JMP}\; l].c) \to (s, \sigma, c')} \; \text{JMP}$$

$$\text{JMPC}_\text{T} \; \frac{\begin{array}{c} \text{top } s = \mathbf{T} \\ s' = \text{pop } s \quad c' = \texttt{findlabel}(p, l) \end{array}}{p \vdash (s, \sigma, [\mathbf{JMPC}\; l].c) \to (s', \sigma, c')} \qquad \frac{\begin{array}{c} \text{top } s = \mathbf{F} \\ s' = \text{pop } s \end{array}}{p \vdash (s, \sigma, [\mathbf{JMPC}\; l].c) \to (s', \sigma, c)} \; \text{JMPC}_\text{F}$$

$$\text{ANDs} \; \frac{\begin{array}{c} (t, t') = \text{top2 } s \\ s' = \text{push } (t \wedge t') \; (\text{pop2 } s) \end{array}}{p \vdash (s, \sigma, [\mathbf{ANDs}].c) \to (s', \sigma, c)} \qquad \frac{\begin{array}{c} t = \text{top } s \wedge op \\ s' = \text{push } t \; (\text{pop } s) \end{array}}{p \vdash (s, \sigma, [\mathbf{AND}\; op].c) \to (s', \sigma, c)} \; \text{AND}$$

$$\text{ADDs} \; \frac{\begin{array}{c} (t, t') = \text{top2 } s \\ s' = \text{push } (t + t') \; (\text{pop2 } s) \end{array}}{p \vdash (s, \sigma, [\mathbf{ADDs}].c) \to (s', \sigma, c)} \qquad \frac{\begin{array}{c} t = \text{top } s + op \\ s' = \text{push } t \; (\text{pop } s) \end{array}}{p \vdash (s, \sigma, [\mathbf{ADD}\; op].c) \to (s', \sigma, c)} \; \text{ADD}$$

$$\text{GTs} \; \frac{\begin{array}{c} (t, t') = \text{top2 } s \\ s' = \text{push } (t < t') \; (\text{pop2 } s) \end{array}}{p \vdash (s, \sigma, [\mathbf{GTs}].c) \to (s', \sigma, c)} \qquad \frac{\begin{array}{c} t = \text{top } s < op \\ s' = \text{push } t \; (\text{pop } s) \end{array}}{p \vdash (s, \sigma, [\mathbf{GT}\; op].c) \to (s', \sigma, c)} \; \text{GT}$$

$$\text{TON-OFF} \; \frac{\text{top } s = \mathbf{F} \quad s' = \text{pop } s \quad \sigma' = \sigma[Tx.Q \mapsto \mathbf{F}, Tx.ET \mapsto 0]}{p \vdash (s, \sigma, [\mathbf{TON}\; Tx, Pt].c) \to (s', \sigma', c)}$$

$$\text{TON-ON} \; \frac{\begin{array}{c} \text{top } s = \mathbf{T} \quad Tx.ET < Pt \\ s' = \text{pop } s \quad \sigma' = \sigma[Tx.Q \mapsto \mathbf{F}, Tx.ET \mapsto Tx.ET + \delta] \end{array}}{p \vdash (s, \sigma, [\mathbf{TON}\; Tx, Pt].c) \to (s', \sigma', c)}$$

$$\text{TON-END} \; \frac{\begin{array}{c} \text{top } s = \mathbf{T} \quad Tx.ET \geq Pt \\ s' = \text{pop } s \quad \sigma' = \sigma[Tx.Q \mapsto \mathbf{T}, Tx.ET \mapsto Tx.ET + \delta] \end{array}}{p \vdash (s, \sigma, [\mathbf{TON}\; Tx, Pt].c) \to (s', \sigma', c)}$$

Figure 5: IL operational semantics

instruction, there is no condition on the branching label value (position of the jumping target) compared to the current position of the program counter. This can lead to non terminating IL programs. In practice this should not be the case, since every IL program should terminate during the scan cycle time limit. We chose here not to consider this kind of errors. They can be treated at the level of the syntactic analysis or by static analysis of the program.

The transition relation for the TON instruction is given by the rules TON-OFF, TON-ON and TON-END of Figure 5. The elapsed time variable $ET$ of the TON timer is incremented by the global constant $\delta$ when the timer is activated (the value of top of the stack is *true*). The timer output $Q$ is activated when the elapsed time variable $ET$ is greater or equal to the timer delay parameter $PT$.

With the inference rules defined above, we have a deterministic operational semantics of the IL language. We can now define a certified translation function from LD to IL.

# 5 Sequential Function Charts

The SFC language is a graphical language for modeling PLC. It is part of the IEC 61131–3 standard and frequently used together with IL and other languages of this standard. SFC are used to describe the overall control flow structure of a system. Due to the graphical nature of the language, we have written a tool which generates Coq representations from graphical SFC models.

The parts of the standard describing SFC leave a few semantical aspects open to the implementation of the PLC modeling and code generation tool. In cases where the semantics is not well defined by the standard we have adapted our semantics aiming to be compatible with the EasyLab [3] tool. The description given in this work follows the description given in [8].

## 5.1 Syntax

Syntactically we represent an SFC as a tuple $(S, S_0, T, A, F, V, \mathit{Val}_V)$. It comprises a set of steps $S$ and a set of transitions $T$ between them. A step is a system location which may either be active or inactive in an actual system state, it can be associated with SFC action blocks from a set $A$. These perform sets of operations and can be regarded as functors that update functions representing memory. Memory is represented by a function from a set of variables $V$ to a set of their possible values $\mathit{Val}_V$. The mapping of steps to sets of action blocks is done by the function $F$.

In our SFC framework, action blocks may be described using the semantics defined in the previous sections. In particular, we have established functions that allow conversion of SFC states into IL states and vice versa. Thus, the execution of an action block realized in IL comprises the following steps:

- Conversion of the SFC state into an IL state
- Execution of the IL program associated with the action block using the semantics from Section 4.
- Update of the SFC state by using the final IL state.

For LD we have established a conversion into IL.

A transition is a tuple $(S_{in}, g, S_{out})$. It features a set of steps that have to be enabled $S_{in} \subseteq S$ in order to take the transition. It features a guard $g$ that has to be evaluated to true for the given system state. The guard $g$ is a function from system memory to a truth value – in Coq we formalize this as a function to the *Prop* datatype. A transition may have multiple successor steps $S_{out} \subseteq S$. The types $\mathit{Val}_V$ that are formalized in our SFC language comprise different integer types and boolean values. The set of SFC steps includes also a set $S_0 \subseteq S$ representing the initially active steps.

Figure 6 shows an example of an SFC structure realizing a loop with a conditional branch. The execution starts with an initialization step *init*. After it has been processed control may pass to either *Step2* or to a step *Return*. Once *Step2* has been processed control is passed to *init* again.

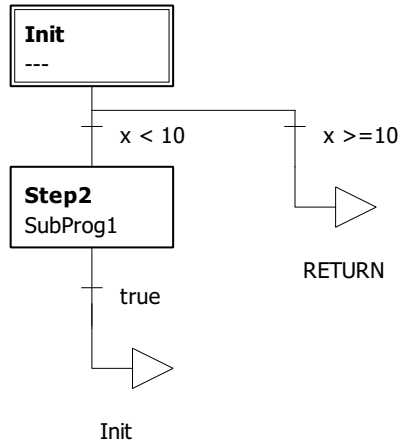

Figure 6: A loop in the SFC language

Please note that in addition to loops and branches, SFC allows also the definition of parallel processing and synchronization of control. This is due to the multiple successor and predecessor steps in a transition.

The Coq realization of the SFC syntax follows the presented description. One goal is compatibility with the SFC implementation of the EasyLab tool. Here, to ease the generation, we distinguish between steps and step identifiers in our Coq files, thereby introducing some level of indirection.

## 5.2 Semantics

Semantically the execution of an SFC encounters states, which are $(m, s, a)$ tuples. They are characterized by a memory state $m$, the function from variables to their values, a set of active steps $s$ and a set of active action blocks $a$ that need to be processed.

The semantics is defined by a state transition system which comprises two kinds of rules:

1. A rule for processing an action block from the set of active action blocks $a$. This corresponds to updating the memory state and removing the processed action block from $a$.

2. A rule for performing a state transition. The effect on the system state is that some steps are deactivated, others are activated. Each transition needs a guard that can be evaluated to true and a set of active steps. Furthermore, we require that all pending action blocks of a step that is to be deactivated have been executed.

It is custom to specify the timing behavior of a step by using a (maximal) execution time associated with it (cycle time). In our work, this is realized using special variables that represent time.

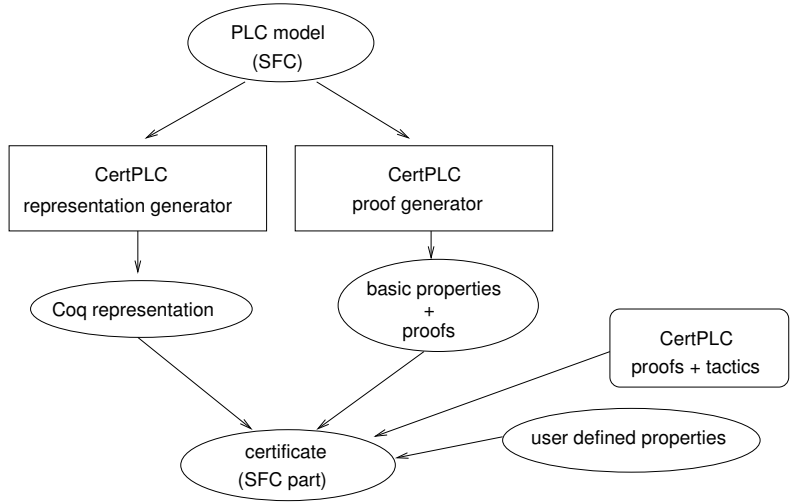

Figure 7: CertPLC overview

# 6 Tool Support for SFC and Proving Principles

For the generation of Coq representation files from the graphical SFC language and reasoning about them we have created a tool+associated Coq files (CertPLC, described in [6, 5]). The tool is implemented in Java and uses SFC files built with a graphical PLC configuration environment: EasyLab [3]. CertPLC software components are implemented as an Eclipse plug-in in Java using the IEC 61131–3 meta model of EasyLab and the Eclipse Modeling Framework (EMF) [15].

The text-based IL code and LD code provided in our Coq based syntax can be imported directly into the generated Coq files which represent the SFC structure. Otherwise, it is possible to translate FBD first into IL or LD code and integrate it manually. In this section we describe our tool's architecture and frequently used principles for proving properties of SFC structures. Since SFC is used to represent the global control flow of on IEC 61131–3 system, proof support for SFC is especially crucial.

Figure 7 shows the CertPLC ingredients and their interconnections. In an invocation of the CertPLC tool an SFC model is given to a

- **representation generator** which generates a Coq representation out of it. This is included in one or several files containing the model specific parts of the semantics of the SFC model. The Coq representation is human readable and can be validated against the original graphical SFC specification by experienced users.

The same SFC model is given to a

- **proof generator** which generates Coq proof scripts that contain lemmas and their proofs for some basic properties that state important facts needed for machine handling of the proofs of more advanced properties.

In order to achieve a proof script of a property that shall be ensured on the SFC structure, one needs to formalize this desired property. The property is proved in Coq by using either a provided tactic or a hand written proof script. This is facilitated by provided tactics. These can use the generated properties and their proofs – provided by the proof generator – and a collection of

- **proofs and tactics**, a kind of library. It contains additional preproved facts and tactics which may be used to automatically prove a class of properties.

Furthermore, in our tool framework additional aspects for our specifications can be stored in a Coq library that can be used by generated and non-generated Coq files. It allows storage of often used definitions in addition to the elements described so far.

## 6.1 The CertPLC Tool Environment and Coq

Here we describe Coq specific parts of the CertPLC tool. We present some static Coq code that is generic to our framework. Furthermore, we present some PLC specific example Coq code – definitions and proofs – to demonstrate aspects of its generation.

Taking the semantics sketch of SFC in Section 5 the semantic representation of the SFC structure is encoded in Coq as a transition system. For each given SFC *SFC* we generate a Coq representation. It specifies a set of reachable states and a transition relation.

In order to prove properties of PLC we need files that contain semantics of systems, interesting properties and proofs of these properties. Some of these files are generic, i.e., they can be used for a large class of PLC, properties, and proofs. Here, the Coq formalizations described in earlier sections serve as generic aspects. Other files are highly specific to distinct PLC. For each PLC CertPLC generates files that are just needed for this particular PLC, properties formulated on it, and proofs that can be conducted on it.

## 6.2 Generic / Static Parts of the Coq Infrastructure

The generic parts of the Coq code in our CertPLC tool framework are realized as static Coq files and can used by the dynamically created files.

**Building Blocks for SFC** Building blocks define common elements for the construction of SFC. They comprise, e.g., commonly used action blocks – if not specified using IL, LD or FBD – or parts of it. Formalization of blocks has to be done together with case studies since different application domains have different sets of possible action blocks. Elements that are highly specific to a single application or an application domain are highly common in PLC. We have experienced even vendor specific elements.

**Generic Semantics Framework for SFC** The Coq realization of the SFC syntax follows the description presented in Section 5. For compatibility with the EasyLab tool and to ease generation we distinguish between steps and step identifiers in our Coq files,

```
Definition executeAction:
   fun c c' =>
      let '(m,activeA,activeS) := c in
      let '(m',activeA',activeS') := c' in
         (exists a, In a activeA /\ m' = a m /\
          activeA' = remove Action_eq_dec a activeA) /\
          activeS = activeS'.
```

Figure 8: The *executeAction* predicate

thereby introducing some level of indirection. Most importantly, our semantics framework comprises a template for a state transition relation of PLC systems and a template for defining the set of reachable states. In order to realize this, we first define generic instantiable predicates that formalize a state transition relation. We provide a predicate *executeAction* defined in Figure 8 to give a look and feel. It formalizes the effect of the execution of an action block: The predicate takes two states (sometimes called configurations $c$ and $c'$) and returns a value of type *Prop*. We require four conditions to hold in order to take a state transition:

1. An action block $a$ needs to be active.

2. The memory mapping after the transition is the application of $a$ to the previous memory mapping. This is the updating of the memory by executing the action block.

3. The action block $a$ is removed from the set of active action blocks during the transition.

4. The rest of the state does not change.

Another predicate *stepTransition* formalizes the effect of a transition from a set of SFC steps to another. Here we require the following items:

1. The validity of the transition (guard expression).

2. The memory state may not change.

3. The activation of steps is conform to the semantics.

4. The activation and requirements of action blocks is semantics conform.

Using these predicates we define inductively the set of reachable states as a predicate. It depends on an initial state (comprising a list of initially active steps), and a transition relation. It is defined following the description in Section 5.

**Structural Tactics for SFC** Our framework supports structural tactics that perform the most basic operations for proofs of properties. They work with semantics definitions based on our generic semantics framework. Depending on the property such a tactic is selected and applied as the first step in order to prove the desired property. Different tactics have to be selected: Selection depends on whether the property is some kind of inductive invariant – the default case mostly addressed in this paper – or another class of properties. We have identified several other classes that are relevant for different application domains. Such a tactic is applied as the first step in order to prove the desired property. These tactics already perform most operations concerning the system structure. Especially for the non-standard cases, tactics applications may leave several subgoals open. These may be handled with more specialized tactics tailored for the corresponding characteristics of these proof-goals.

**Arithmetic tactics** Arithmetics tactics solve subgoals that appear at later stages in the proof. Assuming that the effect of actions, e.g., specified using our IL semantics has been symbolically computed, they may be called by structural tactics or work on open subgoals that are left open by these tactics. They comprise classical decision procedures like (e.g., Omega [21] – its implementation in Coq).

Up till now, we are only using existing tactics designed by others. However, we are also working on tactics that combine arithmetic aspects with other system state dependent information.

### 6.3 Semantics Definitions as State Transition Systems

As seen in Section 6.2 we only need to instantiate a template in order to create a system definition that captures the semantics of our PLC. We need to provide at least a set of initially active steps, a transition relation, and action block definitions.

For the initial step, we provide an initial memory state, where all values are set to a default value and a single active entry step.

The transition relation is generated by translating the SFC transition conditions into Coq. The generated Coq elements of the transition relation for the SFC depicted in Figure 6 are shown in Figure 9. Three tuples are shown, each one comprises a set of activated source steps, a condition and a set of target steps activated after the transition. It can be seen that the condition maps a variable value mapping – part of the SFC state – to a truth condition – returning the type *Prop*.

Appropriate action blocks are selected by their names. In addition, to this, we generate several abstract datatype definitions for identifying steps with names and identifiers and function blocks and action blocks.

### 6.4 Automatically Generated Proofs for System-specific Facts

Apart from generating SFC representations, CertPLC can automatically generate for each system some basic properties and proofs. These prove some system-specific facts

```
( Init::nil ,
  fun m => ((fun  (x : int16) => x <int16 10 )  (m VAR_x) ),
  Step2::nil )

( Step2::nil ,
  fun m => ((fun  (x : int16) => 1 )  (m VAR_x) ),
  Init::nil )

( Init::nil ,
  fun m => ((fun  (x : int16) => x >=int16 10 )  (m VAR_x) ),
  Return::nil )
```

Figure 9: Generated transition rules in Coq

```
Lemma aux_1:
   forall s, SFCreachable_states s -> (forall a, In a (snd s) ->
        (  a = action_Init \/  a = action_Step2 )).
```

Figure 10: An automatically generated basic property

of the system. These proofs are used automatically by tactics, but can also be used manually to prove additional user defined properties of systems.

One important fact that needs to be proven is that only those action blocks may appear in the set of currently active action blocks that do belong to the actual system definition. Our proof generator generates an individual lemma and its proof for each PLC. Figure 10 shows such a lemma for an SFC that comprises just two possible action blocks: *action_Init* and *action_Step2*. The predicate *SFCreachable_states* is created by instantiating a template definition from the generic semantics framework for a concrete PLC. *In* and *snd* (second) are Coq functions to denote membership in a set and select an element of a tuple, respectively. In the case at hand *snd* selects the set of active action blocks from a state. The proof script itself is also generated. It comprises an induction on reachable states of the concrete system. Depending on the number of action blocks in the PLC it can typically comprise several hundred applications of elementary Coq tactics.

### 6.5 Proving Invariant Properties of an SFC

We describe the principles of automatically or interactively proving properties correct. Proof scripts encapsulating these principles are generated by the CertPLC framework components as described in above. We focus on inductive invariants.

#### 6.5.1 Proof Structure for Inductive Properties

We start with an inductive invariant property *I* and an SFC description of a PLC *SFC*. Following the ideas presented in [10] the structure of a proof contained in our Coq files is realized by generated proof scripts, generic lemmas and tactics. They establish a proof principle that proves the following goal:

$$\forall\ s\ .\ s \in \mathit{Reachable}_{SFC} \Longrightarrow I(s)$$

The set of reachable states for $SFC$ is denoted $\mathit{Reachable}_{SFC}$. $[\![SFC]\!]$ specifies the state transition relation. First we perform an induction using the induction rule of the set of reachable states. This rule is automatically established by Coq when defining inductive sets. After the application the following subgoals are left open:

$$I(s_0) \text{ for initial states } s_0 \qquad I(s) \wedge (s, s') \in [\![SFC]\!] \Longrightarrow I(s')$$

The first goal can be solved in the standard case by a simple tactic which checks that all conditions derived from $I$ are fulfilled in the initial states.

For the second goal we realize a proof script which – in order to allow efficient proof checking – performs most importantly the following operations:

- Splitting of conjunctions in invariants into independently verifiable invariants.

- Splitting of disjunctions in invariants into two independently verifiable subgoals.

- Normalizing arithmetic expressions and expressions that make distinctions on active steps in the SFC.

- Exhaustive case distinctions on possible transitions. Each case distinction corresponds to one transition in the control flow graph of the SFC. A typical case on a transition from a partially specified state $s$ to a partially specified succeeding state $s'$ can have the following form:

$$\forall\ s\ s'\ .$$
$\quad I(s)$ and case distinction specific conditions on $s\ \wedge$
$\quad$ case specific transition conditions that need to be true to go from $s$ to $s'\ \wedge$
$\quad$ case distinction specific definition of $s' \quad \Longrightarrow \quad I(s')$

The case distinction specific parts in such a goal can, e.g., feature arithmetic constraints, which can be split into further cases.

Some of the cases that occur can have contradictions in the hypothesis. Consider for example an arithmetic constraint for a variable from a precondition of a state contradicting with a condition on a transition. These contradictions result from the fine granularity of our case distinctions. Some effort can be spent to eliminate contradicting cases as soon as possible (cf. [10]) which can speed up the checking process.

Unlike in classical model-checking we get the abstraction from (possibly infinite) concrete states to (finite) arcs in the control flow graph almost for free. Thus, in our case distinctions, we do not have to regard every possible state, we rather partition states into classes of states and reason about these classes symbolically.

- The final step comprises the derivation of the fact that the invariant holds after the transition from the transition conditions and the decision of possible arithmetic constraints.

[10] features a completeness result for a class of inductive invariants for a similar problem.

### 6.5.2 Proving Non Inductive Invariants for SFC

The main focus of CertPLC is on inductive invariants, However, additionally we have experienced the necessity to prove the (un-)reachability of certain states. In particular proof templates for the following cases turned out to be necessary in our case studies:

- State $s$ can only be reached via a transition where a condition $e$ must be enabled, $s$ is not initial, $e$ can never be true in the system, this implies $s$ can not be reached.

- Under system specific preconditions: Given an expression over states $e$, if $e$ becomes true the succeeding state will always be $s$. This is one of the few non-inductive properties. However, the proof of this benefits from a proof that $e$ can only become true in an explicitly classified set of states. This can be provided by one of the techniques above.

## 6.6 CertPLC and the Trusted Computing Base

One might argue that using CertPLC to generate SFC representations introduces a level of indirection which might be a drawback for certification issues. However, we believe that this is justified due to the graphical nature of the SFC language and the relatively short size of the associated trusted computing base (TCB).

Apart from components like operating system and hardware, in the approach realized by CertPLC, the TCB comprises the proof checker (the core of the Coq theorem prover) and the program that generates formal SFC descriptions for Coq automatically. The check that these descriptions indeed represent the original SFC can be done manually. One goal for the generation is human readability to make such a check feasible at least for experienced users. Not part of the TCB are the proof description and its generation mechanism. The proof description only provides hints to the Coq proof checker. In case of faulty proof descriptions a valid property might not be accepted by the Coq proof checker. It can never occur that a faulty property is accepted even if wrong proof descriptions are used. Thus, our approach is sound, but not necessarily complete.

# 7 Handling Additional Languages

This section features a description of the semantics of FBD and LD. We describe our Coq based formal definition of FBD and LD. Both languages feature graphical elements and share some syntactical aspects. However, LD is for historical reasons more popular and is implemented by almost all PLC systems. It is less general than FBD and can be represented relatively easy in a syntactic way that resembles the graphical representation.

Moreover, its semantics makes LD more suited for formal reasoning. For this reason, we keep the discussion on FBD rather high-level and describe LD in more details. Both languages allow the specification of distinct actions to be performed by the PLC.

## 7.1 Function Block Diagrams

FBD closely resemble electrical wiring of elements. They allow an almost arbitrary wiring of function blocks which perform, e.g., computations, serve as delay elements or allow access to variables. The wiring realizes data exchange between these elements.

### 7.1.1 Syntax of FBD

FBD are composed of function blocks which are associated with ports. These are connected using edges. The idea is that each function block is associated with some functionality and data is exchanged via the ports and their connections. Function blocks and edges build up graphs. In principle these graphs can have almost arbitrary topologies. Thus, it is, e.g., allowed to construct cycles within the graphs. In order to determine the functionality of an FBD one tries to compute a fixed-point within a limited number of time (cycle time). The results computed within this time are then written back to global variables and, e.g., given to the PLC controlled system. We have formalized FBD and their semantics in Coq.

In our formal FBD description function blocks comprise a block identifier $b$ and a set of ports $P_b$. Thus, a function block can be syntactically represented as a tuple $(b, P_b)$. An edge connects two ports of function blocks with each other. Thus, an edge connecting the port $p_{b_1}$ of function block $b_1$ with port $p'_{b_2}$ of function block $b_2$ is represented as a tuple $((b_1, p_{b_1}), (b_2, p'_{b_2}))$.

The entire FBD structure comprises a set of function blocks $B$, a set of edges $E$, and a cycle time $t$, the execution time of the FBD. It is represented as a tuple $(B, E, t)$.

### 7.1.2 Semantics of FBD

For assigning meaning to an FBD structure we have on the one side a global system state. This comprises variables $V$ and their sets of possible values $Val_V$. The global system state is a mapping from variables to these values $V \to Val_V$.

Ports of function blocks are also associated with a set of possible values which may either be read or set. For a function block $b$ with a port $p_v$ the set of possible values is $Val_{(p_b)}$. Each function block is equipped with a function that reads its port values and updates them. This function also takes the global system state into account $f_b$:

$$(V \to Val_V) \to (P_b \to Val_{p_b}) \to (P_b \to Val_{p_b})$$

Furthermore, every function block can be equipped with a function that updates the global system state. It also depends on local port values.

$$(V \to Val_V) \to (P_b \to Val_{p_b}) \to (V \to Val_V)$$

**Semantics as total order of partial orders of function blocks**  The semantics of an FBD can be divided into different time-ticks. Each time tick is associated with a partial order of function blocks: $T_1, ..., T_n$. There is a total order on time-ticks: $T_1 \leq ... \leq T_n$. The partial order on function blocks is denoted $T_i = (B_i, \preceq_i)$. The idea is that we evaluate time-ticks using the global order. Each time tick is evaluated function block wise using the partial order of the time-tick. This can lead to non-determinism.

**Deriving Semantics from the Syntax**  Certain function blocks can be marked with a delay. That is we have a function $d$ that maps function blocks to a time delay. A time-tick contains the maximal number of function blocks that can be evaluated at the current time.

In order to compute the partial order of function blocks we start at time $t = 1$ and include all blocks that can be evaluated in this tick, then we increase it and include the next bunch of function blocks. We do this again until the cycle time is reached. In case of cyclic structures blocks can be included in several time-ticks.

## 7.2 Ladder Diagrams

LD is a widely used graphical language for programming PLC. It is very suitable for writing control functionality. An LD program corresponds to a circuit of relay logic components. Each LD diagram is composed of two vertical lines representing the *power rails*. This *power rails* are connected by horizontal lines called rung. A rung is a circuit representing an operation of the LD program. An LD diagram is read from left to right and from top to bottom (Figure 11). Following this order, the rungs of an LD program are executed until an END rung is reached.

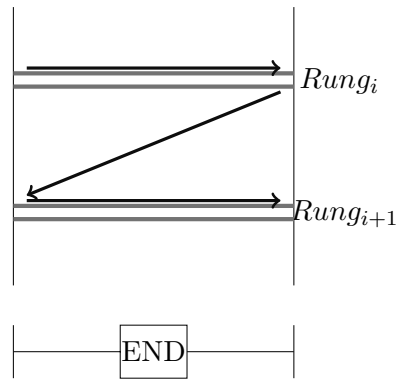

Figure 11: LD structure

Every rung must start with one or more inputs and end with at least one output. Figure 12 gives two simple examples of LD programs with the standard representation of inputs (contacts) and output (coils). The first example contains two inputs: a normally open one $I_0$ and a normally closed one $I_1$. The output $Q_0$ is activated if $I_0$ is activated or $I_1$ is not ($Q_0 = I_0 \vee \overline{I_1}$). The second example shows a rung with two outputs. The first output $Q_0$ will be activated if both inputs $I_0$ and $I_1$ are activated ($Q_0 = I_0 \wedge I_1$). The second output $Q_1$ will be activated if the three inputs $I_0$, $I_1$ and $I_2$ are activated ($Q_1 = I_0 \wedge I_1 \wedge I_2$).

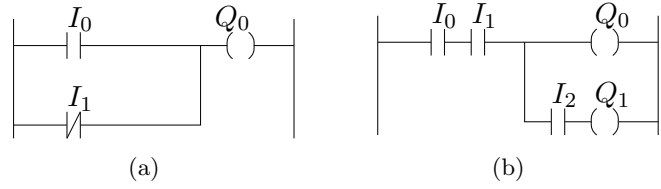

Figure 12: Examples of LD rungs

### 7.2.1 Syntax

We mentioned before that an LD program is a list of rungs. From a syntactic point of view we can decompose a rung into two parts: a *front* and a *rear*.

A *front* corresponds to a logical formula over the rung inputs. An input can be normally open or close. The sequential and parallel connections of the inputs represent respectively conjunction and disjunction.

$$front \quad ::= \quad \texttt{contact}\ id\ |\ \texttt{contactn}\ id\ |\ front\ \&\ front\ |\ front\ ||\ front$$

A rear of a rung is one or more outputs connected in parallel. In a rear, an output can be preceded by a front circuit like in the example of Figure 12(b). An output can be a simple coil (-( )-), a negated coil (-(/)-), a set (latch) coil (-(S)-) or a reset (unlatch) coil (-(R)-).

$$\begin{aligned} output &\ ::=\ \texttt{coil}\ id\ |\ \texttt{coiln}\ id\ |\ \texttt{set}\ id\ |\ \texttt{reset}\ id \\ rear &\ ::=\ output\ |\ front\ \&\ rear\ |\ rear\ ||\ rear \end{aligned}$$

A rung is defined as follow: a front followed by a rear, a conditional jump according to the result of the evaluation of a front, a label or the end rung. Normally a label is just added in front of the rung; but for simplicity we choose to define a special rung for labels. This will make syntactically easier to handle rungs.

$$\begin{aligned} rung &\ ::=\ front\ \times\ rear\ |\ front\ \times\ \texttt{JMPC}\ lb\ |\ \texttt{Label}\ l:\ |\ \texttt{REND} \\ code &\ ::=\ rung_1,\ rung_2, \cdots,\ rung_n \end{aligned}$$

**Model constraints** In our model of LD rungs, a coil cannot be followed by a contact and two coils cannot be connected sequentially like in the example of Figure 13(a). This type of rungs is acceptable from the IEC standard point of view but rarely used in practice. The main reason is that it makes the circuit less readable. The rung in Figure 13(a) can be represented in our model by putting the outputs in parallel lines like in Figure 13(b). The execution of the two rungs gives the same results.

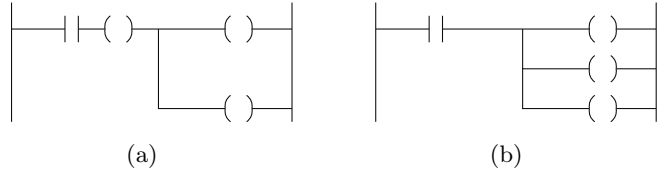

Figure 13: Examples of LD rungs (2)

In LD, functions or subroutines can be defined and called later by the main program. This allows the definition of timers and counters. We do not support this feature of LD but an extension of our model to include function call is possible.

### 7.2.2 Semantics

Using operational semantics, we defined a small-step semantics of LD. This semantics corresponds to a relation over LD configurations. An LD program configuration is a variables state function $\sigma$ and a code sequence $r_1, r_2, \cdots, r_n$. Since we consider only boolean variables, an LD state is a function from variable identifiers to boolean values: $\sigma : Var \to \mathbb{B}$.

The code sequence in the configuration represents the current position in the program where the evaluation will take place. It encodes the rung $r_1$ that will be evaluated in the current transition and its suffixes $r_2, \cdots, r_n$ in the global program. This is equivalent to the *continuation* defined in [2] and used to define some of the semantics of the languages used by the CompCert [18] compiler.

$$\frac{f = \texttt{contact } x \qquad \sigma(x) = b}{f \xRightarrow{\sigma} b} \qquad \frac{f = \texttt{contactn } x \qquad \sigma(x) = b}{f \xRightarrow{\sigma} \neg b}$$

$$\frac{f = f_1 \mathbin{\&} f_2 \qquad f_1 \xRightarrow{\sigma} b_1 \qquad f_2 \xRightarrow{\sigma} b_2}{f \xRightarrow{\sigma} b_1 \wedge b_2} \qquad \frac{f = f_1 \mathbin{||} f_2 \qquad f_1 \xRightarrow{\sigma} b_1 \qquad f_2 \xRightarrow{\sigma} b_2}{f \xRightarrow{\sigma} b_1 \vee b_2}$$

Figure 14: LD operational semantics 1/2: front evaluation

The LD operational semantics is defined by the inference rules given in Figures 14 and 15. The transition relation is decomposed in two parts. The first one is the evaluation of the front of the rung. It is defined by the four inference rules of Figure 14 and denoted $\xRightarrow{\sigma}$ with $\sigma$ a state function.

Figure 15 defines the rung evaluation according to its rear structure. The first four transition rules of Figure 15 correspond to the evaluation of a rear coil: simple, negated, latched or unlatched. The two following inference rules define the inductive evaluation of a rung rear. The rule FREAR corresponds to the case where the rear is composed of a front followed by a rear. In this case, the state transition corresponds to the one where the rung is composed of the conjunction of the two front, the original one and the rear front, and the new rear. The rule ORREAR defines the state transition for the parallel

$$
\text{COIL} \frac{f \stackrel{\sigma}{\Longrightarrow} b \quad \sigma' = \sigma[x \mapsto b]}{p \vdash (\sigma, [f \times \texttt{coil } x].c) \to (\sigma', c)} \quad \frac{f \stackrel{\sigma}{\Longrightarrow} b \quad \sigma' = \sigma[x \mapsto \neg b]}{p \vdash (\sigma, [f \times \texttt{coiln } x].c) \to (\sigma', c)} \text{COILN}
$$

$$
\text{LATCH} \frac{f \stackrel{\sigma}{\Longrightarrow} b \quad \sigma' = \sigma[x \mapsto \sigma(x) \vee b]}{p \vdash (\sigma, [f \times \texttt{set } x].c) \to (\sigma', c)} \quad \frac{f \stackrel{\sigma}{\Longrightarrow} b \quad \sigma' = \sigma[x \mapsto \sigma(x) \wedge \neg b]}{p \vdash (\sigma, [f \times \texttt{reset } x].c) \to (\sigma', c)} \text{UNLATCH}
$$

$$
\text{FREAR} \frac{(\sigma, [f \& f' \times r].c) \to (\sigma', c)}{p \vdash (\sigma, [f \times f' \& r].c) \to (\sigma', c)} \quad \frac{(\sigma, [f \times r].c) \to (\sigma'', c) \quad (\sigma'', [f \times r'].c) \to (\sigma', c)}{p \vdash (\sigma, [f \times r \mathbin{||} r'].c) \to (\sigma', c)} \text{ORREAR}
$$

$$
\text{JMPC}_{\text{FF}} \frac{f \stackrel{\sigma}{\Longrightarrow} false}{p \vdash (\sigma, [f \times \texttt{JMPC } l].c) \to (\sigma, c)} \quad \frac{f \stackrel{\sigma}{\Longrightarrow} true \quad c' = \texttt{findlabel}(p, l)}{p \vdash (\sigma, [f \times \texttt{JMPC } l].c) \to (\sigma, c')} \text{JMPC}_{\text{TT}}
$$

$$
\text{LABEL} \frac{}{p \vdash (\sigma, [\texttt{Label } l :].c) \to (\sigma, c)} \quad \frac{}{p \vdash (\sigma, [\texttt{REND}].c) \to (\sigma, c)} \text{REND}
$$

$$
\frac{}{p \vdash (\sigma, nil) \to (\sigma, nil)} \text{NIL}
$$

Figure 15: LD operational semantics 2/2: rear evaluation

composition of rears. The corresponding transition is the composition of the transition for the rung with first rear followed by the one with the rung with the second rear. In the execution of a rung with parallel rears, the ordering of the rears from top to bottom is important. If one of the inputs of the rung front is updated by the rung top rear, this update will not be taken into account by the evaluation of the rung bottom rear. Also, if the first rear (the top rear) changes a variable that is later modified by the second rear (the bottom rear), only the change made by the last one will be taken into account after the execution of the rung. The situations described above occurs rarely in practice. But we need here to make this choices for the definition of a deterministic formal semantics.

In the branching rung inference rule JMPC$_{\text{TT}}$, we use a function to find the target of the *jump*. For an LD diagram $p$ and a label $l$, $\texttt{findlabel}(p, l)$ returns the maximal suffix of the sequence of rungs $p$ that starts with the label rung $\texttt{Label } l :$. The three last rules correspond respectively to the case of a label rung, a *end of program* rung REND and the empty configuration case. This last rule defines the transition for final or stable configurations. These are the configurations where the code sequence is *nil* or empty. With this rules we have a deterministic operational semantics of the LD language.

**Natural semantics** The big-step semantics of LD corresponds to the transitive closure of the transition relation of the small-step semantics defined above. As we mentioned before, every PLC program should terminate during the cycle scan time. With this termination property, we choose to focus on terminating PLC programs. We define the natural semantics of terminating LD programs as follow:

$$
\frac{p \vdash (\sigma, p) \to^* (\sigma', nil)}{\sigma \stackrel{p}{\Longrightarrow} \sigma'}
$$

In the rule above, a state $\sigma$ is transformed to a state $\sigma'$ after the execution of the program $p$, if the configuration $(\sigma, p)$ transitively evaluates to the final configuration $(\sigma', nil)$.

With the definitions above we have a formal semantics of LD programs. This semantics support a significant subset of the LD language. The next step is the definition of the semantics of the target language of our compiler front-end: the *Instruction List* language.

## 7.3 Transforming LD to IL

Here, we describe a transformation function from LD to IL and investigate its correctness [1]. The translation from LD to IL, is our preferred way of integrating LD code into our framework.

### 7.3.1 Definitions

The translation of an LD program to IL is defined inductively on its list of rungs. Every LD rung is translated to a list of IL instructions. The IL code associated to an LD diagram is the concatenation of the list of IL instructions associated to each of its rungs. Figure 16 gives an example of LD program with the corresponding IL code. The

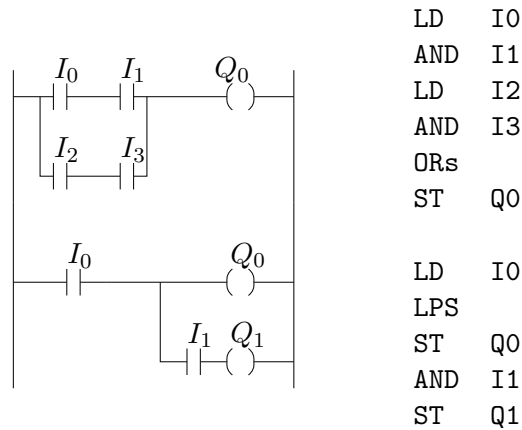

Figure 16: LD to IL: translation example

translation of a rung is done in three parts:

- *front*: a rung front is translated into a list of IL instructions with boolean operators (first five lines of the example of Figure 16).

- *rear*: a rung rear is translated to a single instruction if the rear is a coil or a list of instructions with ending with a store/set operator (last four lines of the example of Figure 16).

---

[1] We do not give here all the details of the formalization. The entire formal definition and proofs are available at the following url: `http://formes.asia/people/sidibiha/plc-coq/`.

- *rung*: the translation of a rung is the concatenation of its front and rear or the corresponding instruction(s) if the rung is a branching or a label or an end one.

Figure 17 gives the Coq definitions of the translation function of a rung front (`front_instrs`) and a rung rear (`rear_instrs`). These function are defined inductively on the structure of the front and the rear. In the first function, if the front is a simple contact the corresponding instruction is a loading one. If the front is the composition (parallel or sequential) of the two front `f1` and `f2` then if the second front `f2` is a simple contact the generated IL instructions will use a simple boolean operator like `OR` otherwise a stack boolean operator like `ORs` will be used. Another method is to use the stack boolean operator in all the cases; but this will generate a more longer program that does not use the optimized operator for the simple boolean operations. The definition of the translation function `rear_instrs` for a rung rear follow the same idea in the case of a rear with a front component. The definition uses the operator `LPS` we introduce in the previous section when generating the IL instructions associated to the parallel composition of two rears. The `LPS` operator will ensure that during the execution the result of the evaluation of the rung front will stay in the memory to be used later in the evaluation of the operations of the second rear.

```
Fixpoint front_instrs (f : RFront) : seq Instr :=
  match f with
    | CONTACT x => [:: LD (var x)]
    | CONTACTN x => [:: LDN (var x)]
    | ROR f1 f2 => match f2 with
                   | CONTACT x => (front_instrs f1) ++ [:: OR (var x)]
                   | CONTACTN x => (front_instrs f1) ++ [:: ORN (var x)]
                   | _ => (front_instrs f1) ++ (front_instrs f2) ++ [:: ORs]
                 end
    | RAND f1 f2 => match f2 with
                    | CONTACT x => (front_instrs f1) ++ [:: AND (var x)]
                    | CONTACTN x => (front_instrs f1) ++ [:: ANDN (var x)]
                    | _ => (front_instrs f1) ++ (front_instrs f2) ++ [:: ANDs]
                  end
  end.
Fixpoint rear_instrs (r : RRear) : seq Instr :=
  match r with
    | rrCOIL r => match r with
                  | COIL x => [:: ST x] | COILN x => [:: STN x]
                  | RSR x => [:: SR x] | RRS x => [:: RS x]
                end
    | rFRear f r' => match f with
                     | CONTACT x => [:: AND (var x)] ++ rear_instrs r'
                     | CONTACTN x => [:: ANDN (var x)] ++ rear_instrs r'
                     | _ => (front_instrs f ++ [:: ANDs]) ++ rear_instrs r'
                   end
    | rORear r1 r2 => [:: LPS] ++ rear_instrs r1 ++ rear_instrs r2
  end.
```

Figure 17: Front and rear translation functions

The rung translation function is defined on top of the front and rear translation functions. The definition is given in Figure 18 (function `rung_instrs`). The LD diagram transformation into IL code is defined by the function `ld2il` of Figure 18. It simply proceed inductively on the list of rungs and concatenate the lists of IL instructions associated to each rung to generate the IL code. With these definitions we have a translation from LD to IL. The next step is to prove the semantic preservation property for this transformation.

```
Definition rung_instrs (r : Rung) : seq Instr :=
  match r with
    | rRung f rr => (front_instrs f) ++ (rear_instrs rr)
    | lRung l => [:: LBL l]
    | rjJMPC f l => (front_instrs f) ++ [:: JMPC l]
    | rEND => [:: RET]
  end.
Fixpoint ld2il (p : Rungs) := if p is r :: p' then rung_instrs r ++ ld2il p' else [::].
```

Figure 18: Rung and LD diagram translation functions

### 7.3.2 Proofs

The proof of the semantics preservation property for the translation from LD to IL corresponds to the following diagram:

$$\begin{array}{ccc} \sigma & \stackrel{p}{\Longrightarrow} & \sigma' \\ \downarrow & & \downarrow \\ (s,\sigma) & \stackrel{ld2il(p)}{\Longrightarrow} & (s,\sigma') \end{array} \qquad (1)$$

The diagram above states that if a state $\sigma$ evaluate to a state $\sigma'$ after the execution of an LD program $p$; then we should have for any stack $s$ that the IL state $(s,\sigma)$ evaluate to the state $(s,\sigma')$ after the execution of $ld2il(p)$, the IL translation of $p$.

We formalized this property in the proof assistant Coq to prove that our translation function is correct. The proof is built in a hierarchical way by proving the correctness of the translation functions for the rung rear and front. These properties are the following:

$$f \stackrel{\sigma}{\Longrightarrow} b \quad \Rightarrow \quad (s, \sigma, [ld2il(f)].c) \to^* (b.s, \sigma, c) \qquad (2)$$

$$\sigma \stackrel{b,r}{\Longrightarrow} \sigma' \quad \Rightarrow \quad (b.s, \sigma, [ld2il(r)].c) \to^* (s, \sigma', c) \qquad (3)$$

The first property states that for any rung front, the evaluation of its translation to IL will terminate with a configuration where the stack will have on top of it a boolean value. This value corresponds to the evaluation of the front according to the state function $s$ (left hand of the implication). The second property states that for any rung rear, the evaluation of its translation to IL starting from a stack $b.s$ [2] will reach a configuration a stack $s$ and a state $\sigma'$. This state $\sigma'$ is the result of the evaluation of the rear starting form the state $s$. The formalization of properties (2) and (3) in Coq corresponds to the two first lemmas of Figure 19.

---

[2] As we mentioned before a rear is always preceded by a front. The $b$ in 3 represents the boolean value associated to this front.

```
Lemma front_instrs_semP : forall d f pc pc' b (st : seq zint) s,
  let p := front_instrs f in eval_rfront_rel s f b ->
    (il_trans d pc !) (st, s%:I, p ++ pc') (b :: st, s%:I, pc').
Lemma rear_instr_state_semP : forall d r pc pc' b (st : seq zint) s s',
  let p := rear_instrs r in eval_rear b r s = s' ->
    (il_trans d pc !) (b :: st, s%:I, p ++ pc') (st, s'%:I, pc').
Lemma ld2il_semP : forall d pc s s' (st : seq zint),
  ld_exec pc s s' -> il_exec d (ld2il pc) (st, s%:I) (st, s'%:I).
```

Figure 19: Semantics preservation proofs

The Lemma `ld2il_semP` of Figure 19 corresponds to the property given in (1). In the statement of the lemma we use a variable `d` to represent the scan cycle duration time. It corresponds to the system parameter $\delta$ we introduce in the definition of the small-step semantics of IL in Figure 5. The propositional predicates `ld_exec` and `il_exec` encode the big-step semantics of the languages LD and IL. They are defined by transitive closure of the small-step semantics of this languages as we mentioned in Section 7.2.2.

# 8 Case Study

Our framework has been applied to verify properties of demonstrators – small-scale machines which are used to validate characterists (also non computer science related properties) of, e.g., a manufacturing chain before a larger plant is set-up – in the industrial automation domain. Figure 20 shows an overview of the SFC structure of a PLC program that controls a sorting station (cf. [9]) on the left side and a picture of the sorting station itself on the right side. Work pieces are transported to two sensors. Based on the values observed by these sensors, a work piece is handled in a different way. The sensor observation is done in the step *workpiece identification*. The handling is done by choosing one of the three alternatives. We have modeled this system in EasyLab and generated the Coq representation of the SFC structure for this case study using Cert-PLC. We have imported the IL programs describing the actions which are taken at the different SFC steps. LD programs may also be used to specify some actions, e.g., Figure 21 shows an LD program that sets some bits for deciding which slide a workpiece has to be sorted in. *is_metallic* and *is_red* are bits set accordingly to sensor values collected from the two sensors. The shown LD code is a part of the step workpiece identification in Figure 20.

Based on this, we have verified that consistency conditions hold. These comprise:

- The verification of inductive invariant based properties. This is described in Section 6.5.1.

- The verification of non-inductive properties. During the conduction of the case study it turned out that non-inductive properties like: Identification of a certain

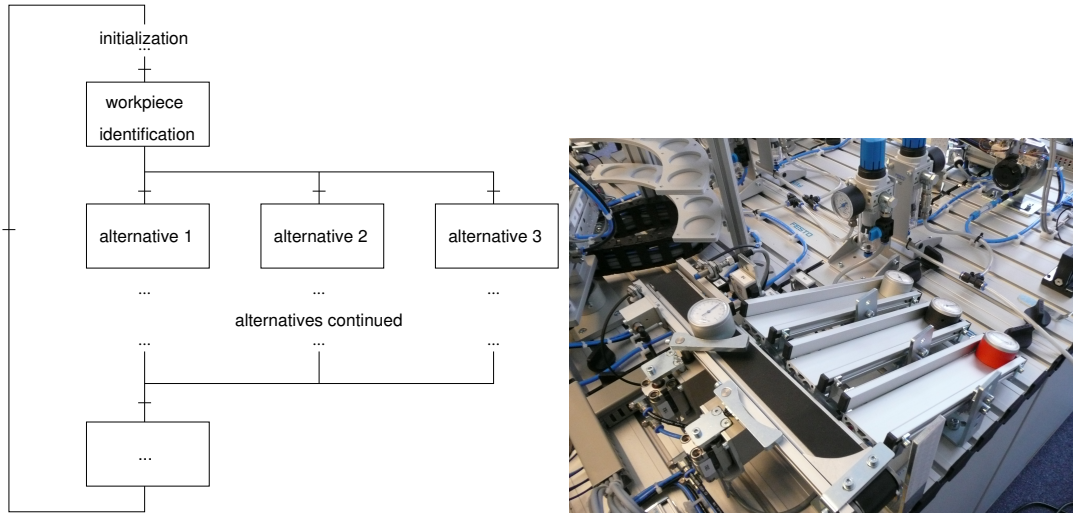

Figure 20: Sorting machine overview

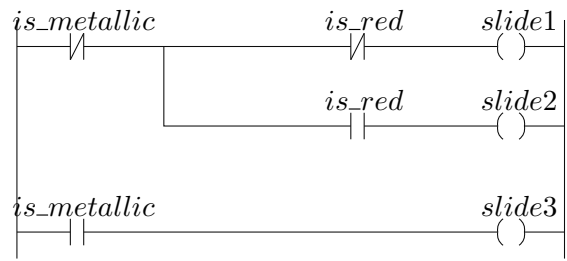

Figure 21: Deciding whether a workpiece shall be sorted to the first, second or third slide

```
    conditions on workpiece
    ->
    let '(m,S,A):= x in
      (conditions on m /\ S = SWorkPieceId::nil /\ A = AWorkPieceId::nil)
    ->
    ...
    transition conditions between x x'
    ...
    ->
    let '(m',S',A'):= x' in
      (some conditions on m' /\ S' = S13::nil /\ A' = nil))
```

Figure 22: Constraint in Coq

work piece implies treatment in a work piece specific way and this occurs within a fixed amount of execution steps, are also of relevance. Mutual exclusion properties of work piece treatment can be proved by doing these work piece specific proofs for all kinds of possible work pieces, first, and using these results for proving the mutual exclusion property.

Proofs for are done in a modular fashion: we verify the effect of IL parts in the PLC execution and use these proofs to derive facts on the execution of several SFC steps.

Figure 22 shows an example of Coq code + pseudo code to give a look and feel on the nature of our proof goals. Given a concrete workpiece and conditions on a state $x$ which corresponds to a state just before the workpiece identification. A succeeding state $x'$ in the SFC language has to fulfill requirements on variable values $m'$ the set of currently active steps $S'$ and currently active actions $A'$ after a certain execution time. In our example it involves several single SFC state transitions (state transition rule applications, cf. Section 5) to get from $x$ to $x'$. This is given in the transition conditions between $x$ and $x'$. The set of currently active steps in the resulting state $x'$ comprises one step $S13$ which corresponds to a step in the first alternative for handling our workpiece.

**Evaluation Aspects:** Coq representation generation for SFC programs and the import of IL and LD code is feasible for IEC 61131–3 based PLC descriptions that are solely described with these languages. Extending the semantics definition for additional commands which may appear in some PLC descriptions is relatively easy, due to the modularity of our semantics framework.

The inductive proof techniques used in the properties generated by the CertPLC tool and the non-inductive proof techniques used manually in the case study have been successfully applied in previous work which did not deal with PLC (e.g., our own work [10]). Here we have demonstrated their applicability for a realistic PLC. Using our Coq semantics and CertPLC, standard properties of a PLC can be verified by experienced

Coq users within several hours. This may result in up to a few hundred lines of proof code for an example as in Figure 22. Common tactic applications are encapsulated into user defined tactics and libraries to further speed this process up, make the scripts smaller, and especially make the approach usable for people who have some knowledge in formal methods but are not Coq experts.

# 9 Related work

Formal treatment of PLC and the IEC 61131–3 standard has been discussed by a larger number of authors before. Formalization work on the semantics of the Sequential Function Charts is given in [11, 12]. This work was a starting point for our formalization of SFC semantics.

Work on the formal treatment of the FBD language – which is also a part of IEC 61131–3 – can be found in [30, 29]. The FBD programs are checked using a model-checking approach.

The approach presented in [25] regards a translation from the IL language to an intermediate representation (SystemC). A SAT instance is generated out of this representation. The correctness of an implementation is guaranteed by equivalence checking with the specification model.

There are plenty of examples of the use of *model checking* for the verification of PLC programs. The paper [4] considers the SFC language. Untimed SFC models are transformed in to the input language of the Cadence SMV tool. Timed SFC models are transformed into timed automata. These can be analyzed by the Uppaal tool. In [19] a semantics of IL is defined using timed automata. The language sub-set contains TON timers but data types are limited to booleans. The formal analysis is performed by the model checker Uppaal.

In [13] an operational semantics of IL is defined. A significant sub-set of IL is supported by this semantics, but it does not include timer instructions. The semantics is encoded in the input language of the model checker Cadence SMV and linear temporal logic (LTL) is used to specify properties of PLC programs.

In contrast to the model checking work, we are using a higher-order theorem prover for our work. In general higher-order theorem provers require a higher level of interaction (we are aiming at overcoming this drawback by generating proof scripts and providing automatic tactics). On the plus side they allow in general richer specifications, abstractions and proofs. In the theorem proving community, there has been some work on the formal analysis of PLC programs. In [26] the theorem prover HOL is used to verify PLC programs written in FBD, SFC and ST languages. In this work, modular verification is used for compositional correctness and safety proofs of programs. For the Coq system, an example of verification of a PLC program with timers is presented in [27]. A quiz machine program is used as an example in this work, but no generic model of PLC programs is formalized. There is also a formalization of a semantics[3] of the LD languages in Coq. This semantics support a sub-set of LD that contains branching instructions. This work is a component of a development environment for PLC.

---

[3]Research report in Korean available at: http://pllab.kut.ac.kr/tr/2009/ldsemantics.pdf

As for CertPLC, notable milestones on frameworks to certify properties of systems comprise proof carrying code [20]. Proofs for program-specific properties are generated during the compilation of these programs. These are used to certify that these properties do indeed hold for the generated code. Thus, users can execute the certified code and have, e.g., some safety guarantees. At least two problems have been identified:

1. Properties have to be formalized with respect to some kind of semantics. This is sometimes just implicitly defined.

2. Proof checkers can grow to a large size. Nevertheless, they have to be trusted.

The problem of trustable proof checkers is addressed in foundational proof carrying code [1, 28]. Here the trusted computing base is reduced by using relatively small proof checkers. The problem of providing a proof carrying code approach with respect to a mathematically founded semantics is addressed in [24]. In previous work we have also addressed the problem of establishing a formal semantics for related scenarios [7, 10].

## 10 Conclusions and Future Works

Programmable Logic Controller applications can be critical in a safety or economical cost sense. Therefore formal verification of PLC programs does increase the confidence in such applications. In this paper we presented a formal framework for the verification of PLC programs written in 4 of the 5 standard languages for programming PLC systems. In addition to the creation of a representation and a proof for a property of a system, the attempts to create proofs of properties can provide valuable feedback during the development of a PLC. The languages are LD, IL, SFC and to some extend FBD. We defined a formal semantics of these languages in the formal proof system Coq. We presented also a certified transformation from LD to IL. This work can be used with the CertPLC tool that automatically generates an SFC formal representation from a graphical representation and some facts + their proofs. Together with CertPLC we reviewed some proving principles and evaluated our framework by proving safety properties for a PLC based real industrial demonstrator.

**Future Work**

A long term goal is the extension the tool support for verification of properties based on the presented framework. Based on this the conduction of larger case-studies is also a goal.

Another perspective of this work is the development of a certified compiler for PLC. This is an ongoing work and we plan to formalize and certify a transformation of PLC programs written in the IL language to C using the work done by the CompCert C certified compiler [18]. An integration of our formal semantics of PLC and the certified compiler to the EasyLab framework is also an interesting perspective. This can lead to a complete environment for the development of certified PLC programs.

# References

[1] A. W. Appel. Foundational proof-carrying code *Logic in Computer Science*. IEEE Computer Society, 2001. (LICS'01).

[2] A. W. Appel and S. Blazy. Separation Logic for Small-step Cminor. Theorem Proving in Higher Order Logics, Kaiserslautern (Germany), January 2007, pp.5-21, vol 4732 of LNCS, Springer-Verlag, 2007. (TPHOLs'07 )

[3] S. Barner, M. Geisinger, Ch. Buckl, and A. Knoll. EasyLab: Model-based development of software for mechatronic systems. Mechatronic and Embedded Systems and Applications, IEEE/ASME, October 2008.

[4] N. Bauer, S. Engell, R. Huuck, B. Lukoschus, and O. Stursberg. Verification of plc programs given as sequential function charts. In *In: Integration of Software Specification Techniques for Applications in Eng., Springer, LNCS*, pages 517–540, 2004.

[5] J. O. Blech. A Tool for the Certification of PLCs based on a Coq Semantics for Sequential Function Charts. http://arxiv.org/abs/1102.3529, 2011.

[6] J. O. Blech. A Tool for the Certification of Sequential Function Chart based System Specifications. 6th International Workshop on Systems Software Verification. Nijmegen, The Netherlands, August 2011.

[7] J. O. Blech and B. Grégoire. Certifying Compilers Using Higher Order Theorem Provers as Certificate Checkers. Formal Methods in System Design, Springer-Verlag, 2010.

[8] J. O. Blech, A. Hattendorf, J. Huang. An Invariant Preserving Transformation for PLC Models. IEEE International Workshop on Model-Based Engineering for Real-Time Embedded Systems Design, 2011.

[9] J. O. Blech and S. Ould Biha. Verification of PLC Properties Based on Formal Semantics in Coq. 9th International Conference on Software Engineering and Formal Methods, Montevideo, Uruguay, 2011. (SEFM 2011)

[10] J. O. Blech and M. Périn. Generating Invariant-based Certificates for Embedded Systems. ACM Transactions on Embedded Computing Systems (TECS). *accepted*

[11] S. Bornot, R. Huuck, Y. Lakhnech, B. Lukoschus. An Abstract Model for Sequential Function Charts. Discrete Event Systems: Analysis and Control, Workshop on Discrete Event Systems, 2000.

[12] S. Bornot, R. Huuck, Y. Lakhnech, B. Lukoschus. Verification of Sequential Function Charts using SMV. Parallel and Distributed Processing Techniques and Applications. CSREA Press, June 2000. (PDPTA 2000)

[13] G. Canet and S. Couffin and J.J. Lesage and A. Petit and P. Schnoebelen. Towards the automatic verification of PLC programs written in Instruction List. In *IEEE International Conference on Systems, Man, and Cybernetics*, 2000.

[14] The Coq Development Team. The Coq System. *http://coq.inria.fr*.

[15] The Eclipse Modeling Framework, *http://www.eclipse.org/modeling/emf/*.

[16] G. Gonthier and A. Mahboubi. A small scale reflection extension for the Coq system. *INRIA Technical report, http://hal.inria.fr/inria-00258384*.

[17] R. Huuck. Semantics and Analysis of Instruction List Programs. In *Electr. Notes Theor. Comput. Sci.*, 2005.

[18] X. Leroy A formally verified compiler back-end. In *Journal of Automated Reasoning*, Vol.43, No.4, pp.363-446, 2009.

[19] A. Mader and H. Wupper. Timed Automaton Models for Simple Programmable Logic Controllers. In *Euromicro Conference on Real-Time Systems*, 1999.

[20] G. C. Necula. Proof-carrying code. *Principles of Programming Languages*. ACM Press, 1997. (POPL'97).

[21] W. Pugh. The Omega test: a fast and practical integer programming algorithm for dependence analysis. *ACM/IEEE Conference on Supercomputing*, pages 4–13, 1991. (SC'91).

[22] S. Ould Biha. A formal semantics of PLC programs in Coq. In *35th IEEE Computer Software and Applications Conference*, Munich, 2011. (COMPSAC 2011)

[23] Programmable controllers - Part 3: Programming languages, IEC 61131-3: 1993, International Electrotechnical Commission, 1993.

[24] R. R. Schneck and G. C. Necula. A Gradual Approach to a More Trustworthy, Yet Scalable, Proof-Carrying Code. *Conference on Automated Deduction*, vol. 2392 of *LNCS*, Springer-Verlag, 2002. (CADE'02).

[25] A. Sülflow and R. Drechsler. Verification of plc programs using formal proof techniques. In *Formal Methods for Automation and Safety in Railway and Automotive Systems*, pages 43–50, Budapest, 2008. (FORMS/FORMAT 2008)

[26] N. Volker and B.J. Kramer, Automated verification of function block-based industrial control systems. In *Science of Computer Programming*, volume 42, pages 101–113, 2002.

[27] H. Wan and G. Chen and X. Song and M. Gu. Formalization and Verification of PLC Timers in Coq. In *33rd IEEE Computer Software and Applications Conference*, 2009. (COMPSAC 2009)

[28] D. Wu, A. W. Appel, and A. Stump. Foundational proof checkers with small witnesses. *ACM Conference on Principles and Practice of Declarative Programming*. ACM Press, 2003. (PPDP'03).

[29] J. Yoo, S. Cha, and E. Jee. A verification framework for fbd based software in nuclear power plants. In *15th Asia Pacific Software Engineering Conference*, Beijing, China, Dec. 3 5, 2008. (APSEC'08)

[30] J. Yoo, S. Cha, and E. Jee. Verification of plc programs written in fbd with vis. In *Nuclear Engineering and Technology*, Vol.41, No.1, pp.79-90, February 2009.